\documentclass[aps,pra,twocolumn,superscriptaddress,showpacs]{revtex4-1}
\usepackage{bm}
\usepackage[dvips]{graphicx}
\usepackage{amsmath,amssymb}
\usepackage{epsfig}
\usepackage{color}
\usepackage{dcolumn}

\newcommand{\ketg}{\mbox{$|g\rangle $}}
\newcommand{\ketk}{\mbox{$|k\rangle $}}
\newcommand{\vk}{\mbox{$\vec k $}}
\usepackage{comment}

\begin{document}
\title{Spectrally resolved detection in transient-reflectivity measurements of coherent optical phonons in diamond} 

\author{Kazutaka G. Nakamura}
\email{nakamura@msl.titech.ac.jp}
\affiliation{Materials and Structures Laboratory, Tokyo Institute of Technology, 4259 Nagatsuta, Yokohama 226-8503, Japan}
\affiliation{Department of Innovative and Engineered Materials, Tokyo Institute of Technology, 4259 Nagatsuta, Yokohama 226-8503, Japan}
\affiliation{CREST, Japan Science and Technology Agency, 4-1-8 Kawaguchi, Saitama 332-0012, Japan}

\author{Kazuma Ohya}
\affiliation{Materials and Structures Laboratory, Tokyo Institute of Technology, 4259 Nagatsuta, Yokohama 226-8503, Japan}
\affiliation{Department of Innovative and Engineered Materials, Tokyo Institute of Technology, 4259 Nagatsuta, Yokohama 226-8503, Japan}
\affiliation{CREST, Japan Science and Technology Agency, 4-1-8 Kawaguchi, Saitama 332-0012, Japan}

\author{Hiroshi Takahashi}
\affiliation{Department of Physics, Faculty of Science and Technology, Tokyo University of Science, 2641 Yamazaki, Noda, Chiba 278-8510, Japan}

\author{Tetsuya Tsuruta}
\affiliation{Materials and Structures Laboratory, Tokyo Institute of Technology, 4259 Nagatsuta, Yokohama 226-8503, Japan}
\affiliation{Department of Innovative and Engineered Materials, Tokyo Institute of Technology, 4259 Nagatsuta, Yokohama 226-8503, Japan}
\affiliation{CREST, Japan Science and Technology Agency, 4-1-8 Kawaguchi, Saitama 332-0012, Japan}

\author{Hiroya Sasaki}
\affiliation{Materials and Structures Laboratory, Tokyo Institute of Technology, 4259 Nagatsuta, Yokohama 226-8503, Japan}
\affiliation{Department of Innovative and Engineered Materials, Tokyo Institute of Technology, 4259 Nagatsuta, Yokohama 226-8503, Japan}
\affiliation{CREST, Japan Science and Technology Agency, 4-1-8 Kawaguchi, Saitama 332-0012, Japan}

\author{Shin-ichi Uozumi}
\affiliation{Materials and Structures Laboratory, Tokyo Institute of Technology, 4259 Nagatsuta, Yokohama 226-8503, Japan}
\affiliation{Department of Innovative and Engineered Materials, Tokyo Institute of Technology, 4259 Nagatsuta, Yokohama 226-8503, Japan}
\affiliation{CREST, Japan Science and Technology Agency, 4-1-8 Kawaguchi, Saitama 332-0012, Japan}

\author{Katsura Norimatsu}
\affiliation{Materials and Structures Laboratory, Tokyo Institute of Technology, 4259 Nagatsuta, Yokohama 226-8503, Japan}
\affiliation{Department of Innovative and Engineered Materials, Tokyo Institute of Technology, 4259 Nagatsuta, Yokohama 226-8503, Japan}
\affiliation{CREST, Japan Science and Technology Agency, 4-1-8 Kawaguchi, Saitama 332-0012, Japan}

\author{Masahiro Kitajima}
\affiliation{Materials and Structures Laboratory, Tokyo Institute of Technology, 4259 Nagatsuta, Yokohama 226-8503, Japan}
\affiliation{CREST, Japan Science and Technology Agency, 4-1-8 Kawaguchi, Saitama 332-0012, Japan}

\author{Yutaka Shikano}
\email{yshikano@ims.ac.jp}
\affiliation{Materials and Structures Laboratory, Tokyo Institute of Technology, 4259 Nagatsuta, Yokohama 226-8503, Japan}
\affiliation{Research Center of Integrative Molecular Systems (CIMoS), Institute for Molecular Science, National Institutes of Natural Sciences, 38 Nishigo-Naka, Myodaiji, Okazaki, Aichi 444-8585, Japan}
\affiliation{Institute for Quantum Studies, Chapman University, 1 University Dr., Orange, California 92866, USA}

\author{Yosuke Kayanuma}
\email{kayanuma.y.aa@m.titech.ac.jp}
\affiliation{Materials and Structures Laboratory, Tokyo Institute of Technology, 4259 Nagatsuta, Yokohama 226-8503, Japan}
\affiliation{CREST, Japan Science and Technology Agency, 4-1-8 Kawaguchi, Saitama 332-0012, Japan}
\affiliation{Graduate School of Sciences, Osaka Prefecture University, 1-1 Gakuen-cho, Sakai, Osaka, 599-8531 Japan}
\date{\today}

\begin{abstract}
Coherent optical phonons in bulk solid system play a crucial role in understanding and designing light-matter interactions 
and can be detected by the transient-reflectivity measurement. In this paper, we demonstrate spectrally resolved 
detection of coherent optical phonons in diamond from ultrashort infrared pump-probe measurements using optical band-pass filters. 
We show that this enhances the sensitivity approximately $35$ times in measuring the coherent oscillations in the transient 
reflectivity compared with the commonly used spectrally integrated measurement. To explain this observation, we discuss its mechanism.
\end{abstract}
\maketitle

\section{Introduction}
Ultrashort optical pulses generate the coherent oscillation of the lattice, which modulates the macroscopic electric susceptibility.
These lattice oscillations are referred to as coherent phonons and can be detected with another ultrashort pulse via intensity modulations in reflectivity or transmissivity~\cite{Dekorsy2000, Nelson1985}. 
Using coherent phonons and a pump-probe type optical measurement, we can directly observe the oscillation of the phonons and measure their dynamics for a wide variety of materials such as semimetals \cite{Cheng1991, Zeiger1992, DeCamp2001, Katsuki2013}, semiconductors \cite{Cho1990, Dekorsy1993, Merlin1996, Misochko2000, Hase2003, Hayashi2014}, superconductors \cite{Chwalek1991, Albrecht1991, Takahashi2011} and topological insulators \cite{Wu2008, Kamaraju2010, Norimatsu2014, Misochko2015}. 
In addition, the coherent phonons in carbon materials, e.g., graphite \cite{Ishioka2008GR}, graphene \cite{Katayama2013, Kim2013}, and carbon nanotubes \cite{Kato2008, Gambetta2006, Sanders2009, Makino2009, Kim2009}, have attracted much attention in studies of electron-phonon coupling.

To excite and measure the coherent phonons, the pulse duration of the pump and probe pulse needs to be shorter than the vibrational period of the phonons. This requirement corresponds to a spectral width of the optical pulse that is much bigger than the phonon energy. The coherent phonon dynamics can be observed as a change in transient reflectivity. While this change depends on the probe light frequency, the reflected light can be measured without needing spectrally resolved methods. This is commonly used and is referred to here as the spectrally integrated detection. In contrast, the spectrally resolved detection shows that the associated change in transient reflectivity at shorter and longer wavelengths oscillates $180^\circ$ out of phase~\cite{Merlin1997,Misochko2004,Mizoguchi2013}. This implies that the sensitivity of the spectrally resolved detection is much higher than that of the spectrally integrated detection because the shorter- and longer- wavelength components cancel each out.

In this paper, we investigate enhancement of the detection sensitivity for the coherent optical phonons using the spectrally resolved detection and found a strong enhancement of approximately $35$ times in diamond. Diamond has a wide band gap (a direct gap of $7.3$~eV and an indirect gap of $5.5$~eV) and a high optical-phonon frequency ($40$ THz). Raman spectroscopy has been widely used to extract properties of diamond because the phonon spectrum is sensitively dependent on the crystal structure of carbon materials such as diamond, graphite, graphene, and carbon nanotubes. Recently, as an application of quantum memory, the dynamics of longitudinal optical (LO) phonons in diamond has been studied~\cite{Lee2010,Lee2011,Lee2011B, England2, England}. On the other hand, the coherent-phonon measurement using ultrashort visible pulses (photon energy of $3.14$~eV) and spectrally integrated detection have recorded $40$-THz coherent optical phonons in diamond and evaluated these lifetimes based  on the difference in impurity rates~\cite{Ishioka2006}. In the present experiment, we used an ultrashort infrared pulse with a central wavelength of  $1.52$~eV and neglected linear and multiphoton absorption effects. 

\section{Spectrally resolved detection in an ultrashort-pulse pump-probe experiment}
\begin{figure}[thb]
\centering
\includegraphics[width=8.5 cm]{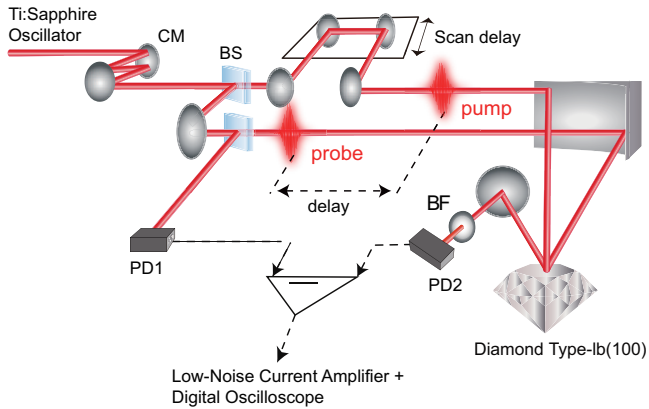}
\caption{(Color online) Schematic of our experimental setup for spectrally resolved detection on the transient-reflectivity measurement. PD denotes the photo-diode measuring pulse intensity, BS denotes a beam splitter, CM is a chirp mirror, and BF is a band-pass filter for spectral cutting.}
\label{expfig}
\end{figure}
\begin{figure}
\centering
\includegraphics[width=6cm]{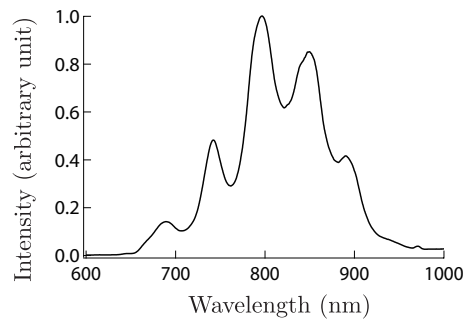}
\caption{Spectrum of the sub-$10$-fs laser pulse. After the experimental procedures, this spectrum was measured at the output port of the Ti:sapphire oscillator with a fiber-type spectrometer.}
\label{spectrumexp}
\end{figure}

The coherent optical phonons are investigated using a pump-probe-type transient reflection measurement (Fig.~\ref{expfig}).  
The output pulse from the Ti:sapphire oscillator (FEMTOLASERs: Rainbow), the spectrum of which was is given in Fig.~\ref{spectrumexp} measured using a USB spectrometer (OceanOptics: USB2000+), was divided into two pulses by a $75/25$ beam splitter, and used as pump and probe pulses. The pump pulse went through a scan delay unit (APE: Scan Delay 50) to control the time delay between the pump and probe pulses. The scan delay was run with a sine function of $20$ Hz. Also, the probe pulse was picked up by a $95/5$ beam splitter to measure the reference beam intensity at a photodiode (PD1). Thereafter, both pump and probe pulses were focused on the sample by using an off-axis parabolic mirror with a focal length of $50$~mm. The reflected pulse from the sample was detected with a photodiode (PD2). In addition, optical bandpass filters are put before PD2 for the spectrally resolved detection, which were FB740, FB800, FB850, and FB900 (Thorlabs Inc.) with transmission at a central wavelength of $740$, $800$, $850$, and $900$~nm, respectively, with a band width of $10$~nm. By applying the opposite bias voltages to PD1 and PD2, we set the balanced detection before the experiment. Its differential signal, to be amplified with a low-noise current amplifier (Stanford Research Systems: SR570), was measured by a digital oscilloscope (Iwatsu: DS5534). To reduce the statistical error, the $32000$ signals were averaged and taken as the measured value. By converting the temporal motion of the scan delay unit to the pump-probe pulse duration, the temporal evolution of the reflectivity change $\Delta R(t)/R_0$ was obtained. 
The time interval of the sampling data points was estimated to be $0.7$~fs. 
Note that the spectral chirping by the optics was compensated using a pair of chirp mirrors in order to minimize  the pulse width at the sample position.
The ultrafast laser conditions in the following experiment were the spectral centroid of $818$~nm evaluated from Fig.~\ref{spectrumexp}, the pulse width $8.9$~fs from the frequency resolved auto correlation measurement (FEMTOLASERS: Femtometer), and the powers $20$~mW and $14$~mW of the pump and probe pulses, respectively. The power and polarization of both the pump and probe pulses were controlled using a half-wave plate and a polarizer.

The sample was a commercially available type-Ib diamond crystal of a $[100]$ crystal plane (SUMITOMO Co.) with a rectangular parallelepiped shape of face size with a $5$~mm $\times$~$5$~mm, and a thickness of $2$~mm. The polarization of the pump pulse was set parallel to the $[110]$ axis. The polarizations of the pump and probe pulses are orthogonal to each other.
\section{Coherent optical phonon dynamics in diamond}
\begin{figure}
\centering
\includegraphics[width=9 cm]{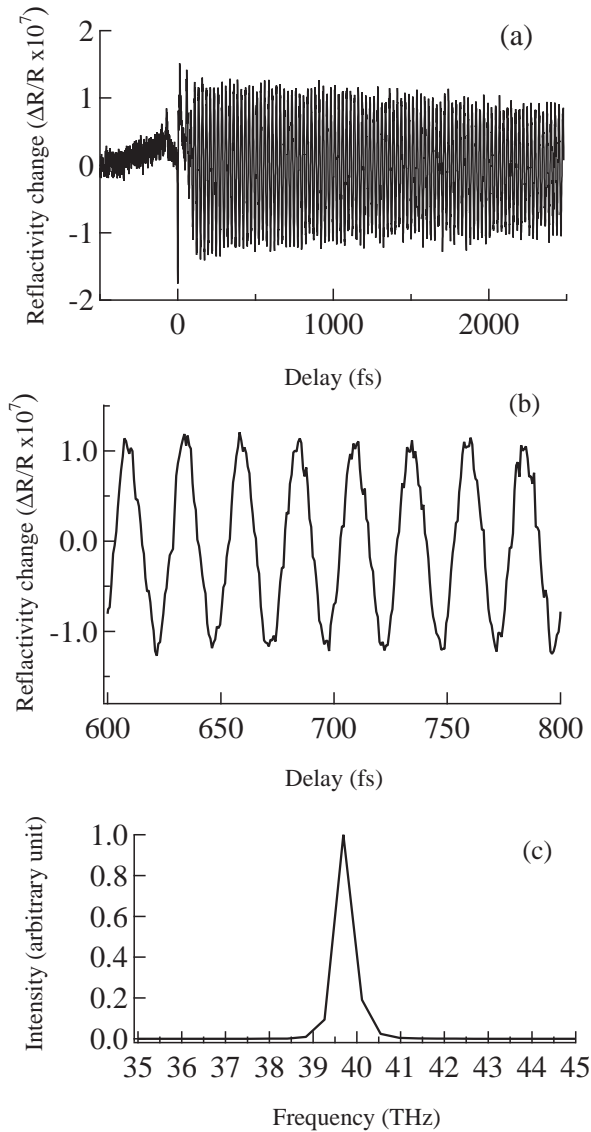}
\caption{Transient reflectivity dynamics from diamond obtained without any filters. (a) The time evolution of a change in reflectivity, (b)  the horizontally expanded signal of the data (a) in a range between $600$~fs and $800$~fs, and (c) the Fourier spectrum of the data (a) in a range between 100 fs and 2400fs.}
\label{typ}
\end{figure}

We next analyze the coherent LO phonon dynamics by measuring the change in transient reflectivity. Figure \ref{typ} shows the transient-reflectivity dynamics of the diamond obtained for a whole spectral range without any filters. With overlapping the pump and probe pulses there is a strong peak at zero delay because of diamond's nonlinear response. After the strong peaks there is a modulation caused by the coherent optical phonons in diamond. The transient reflectivity expanded over the interval from $600$ to $800$~fs clearly indicates that the vibrational period is $25.0 \pm 0.4$~fs (frequency of $40.0 \pm 0.6$ THz). 
The lifetime of this coherent oscillation was estimated to be $6.0 \pm 1.1$~ps  by fitting the data with the damping oscillation. This agrees well with a previous result ($5.59 \pm 1.12$~ps)~\cite{Ishioka2006}. 

\begin{figure}
\centering
\includegraphics[width=8.5 cm]{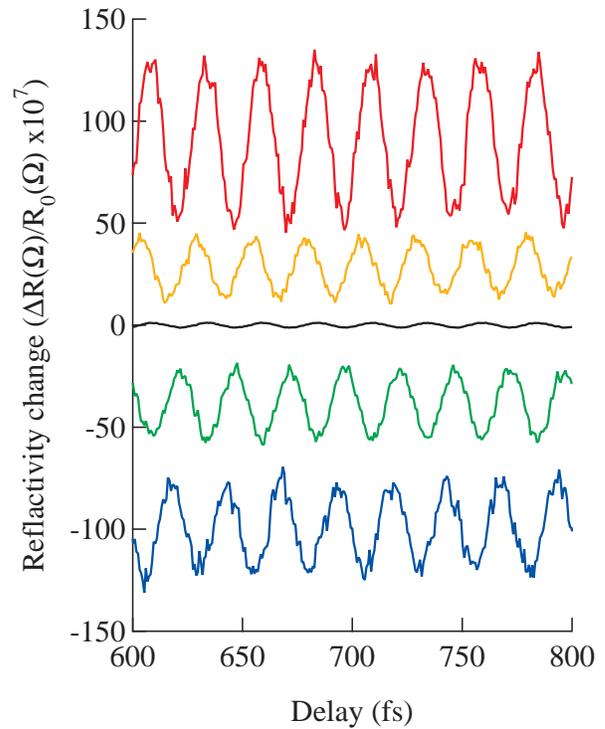}
\caption{(Color online) Transient reflectivity change $\Delta R(\Omega)/R_0 (\Omega)$ of the type Ib diamond obtained without any filters (black) and with filters FB900 (red), FB850 (orange), FB800 (green), and FB740 (blue). The data are plotted with an offset.}
\label{filters}
\end{figure}

Figure \ref{filters} shows the spectrally resolved transient-reflectivity change $\Delta R(\Omega)/R_0(\Omega)$ with bandpass filters FB720, FB800, FB850, and FB900. $R_0(\Omega)$ is the reflectivity measured with a filter without the pump pulse. The amplitude of the coherent oscillations in the transient reflectivity obtained with the bandpass filter is larger than that without the band-pass filter, although the light intensity obtained with the band pass filter is reduced. According to the estimation of the current intensity at PD2, the ratio of the reflected probe pulse intensity with BF900, BF850, BF800, and BF740 compared with that without filtering is estimated to be $0.02$, $0.04$, $0.05$, and $0.02$, respectively. We demonstrate that the oscillation amplitude in $\Delta R(\Omega)/R_0(\Omega)$ with FB900 is $35$ times larger than that without the filter $\Delta R/R_0$.  

The phase of the oscillation on the longer-wavelength side ($850$~nm and $900$~nm) is almost opposite to that on the shorter-wavelength side ($740$~nm and $800$~nm). This phase change in the transient transmission spectrum of carbon nanotubes was already reported~\cite{Kim2013}. The phase of the coherent phonon oscillation in the differential transmission for the radial breathing mode of single-walled carbon nanotubes depends on the detection wavelength. The oscillations at wavelengths of $810$~nm and $780$~nm are anti-phase to each other, and hence below and above the resonance of the $E_{22}$ band gap. In contrast, in the present study, we observed a phase difference in the transient reflectivity although all the probe wavelengths are below the band gap.

To elaborate a mechanism for our observed anti-phase effect, when a delay $\tau_p$ between the pump and probe pulses is much longer than the optical pulse width $\sigma$, the generation and the detection processes are treated as independent. 
To clarify our observation in the spectrally resolved measurement, the detection process of the coherent phonons is only discussed. 
The incidence of the pump pulse on the diamond at time $t=0$ induces coherent optical phonons via impulsive stimulated Raman scattering~\cite{Nelson1985, Merlin1997,Nakamura2015} because the optical energy ($1.52$~eV) is well below the band gap ($7.3$~eV) \cite{Milden2013} of the diamond. 
The induced phonons modulate the polarization $P(t)$ due to the probe pulse $E(t-\tau_p)$ irradiated at $t=\tau_p$ 
as $P(t)=P_0+\Delta P(t)$. Here, $\Delta P(t)$ is proportioainal to the phonon coordinate, $Q(t)=Q_0\cos\omega t$. 
It is shown \cite{Stock1992} that the spectrally resolved reflection modulation $\Delta R(\Omega)$ is given by 
\begin{equation}
\Delta R(\Omega)=2\Omega_0 \mathrm{Im} \left\{E^*(\Omega)\Delta P(\Omega)\right\}, \label{DR}
\end{equation}
in which $E(\Omega)$ and $\Delta P(\Omega)$ are the Fourier components of $E(t-\tau_p)$ and $\Delta P(t)$, and 
$\Omega_0$ is the central frequency of the probe pulse. Equation (\ref{DR}) suggests that in 
$\Delta R(\Omega)$, the optical frequency $\Omega$ and the phonon frequency $\omega$ are synthesized into 
components of frequencies $\Omega\pm\omega$. A quantum mechanical theory yields the formula 
for $\Delta R(\Omega)$ as 
\begin{eqnarray}
\Delta R(\Omega)&=&2\Omega_0\alpha\sqrt{\omega/2\hbar}Q_0
 [ \left\{\chi(\Omega+\omega)-\chi(\Omega\right\}E_0 (\Omega+\omega)
\nonumber\\
&-&\left\{\chi(\Omega)-\chi(\Omega-\omega)\right\}E_0(\Omega-\omega)
]  E_0(\Omega)\nonumber\\
&\times & \sin\omega\tau_p,
\label{final1}
\end{eqnarray}
where $\alpha$ is the dimensionless electron-phonon coupling constant, $\chi(\Omega)$ is 
the electric susceptibility of the crystal, and $E_0(\Omega)\equiv e^{-i\Omega\tau_p}E(\Omega)$ 
is a real quantity in the Fourier transform limit. See technical details in Appendix \ref{appendixA}. 
In the present experiment, the peak position of the spectrum is $800$ nm ($\sim 375$ THz), which is slightly smaller than the spectral centroid ($818$ nm),  and the frequency of the optical phonons is $40$ THz. 
Then, Eq. (\ref{final1}) implies that the largest signals should be measured at $\Omega + \omega = 415$ THz and $\Omega - \omega = 335$ THz, corresponding to $\lambda = 723$ nm and  $\lambda = 895$ nm, 
respectively, which is consistent with the present data.
It is worth noting that Eq. (\ref{final1}) is a universal formula for the spectrally resolved detection of coherent phonons in the transparent region. Note that the prerequisite for the $180^\circ$ out-of-phase effect is that 
the electric susceptibility $\chi(\Omega)$ is a real quantity in the transparent region. In the opaque region, the relative phase of the oscillation depends on the relative magnitude of the real and imaginary parts of  $\chi(\Omega)$.

This formula clearly indicates that the change of the reflectivity at frequency $\Omega\pm\omega$ oscillates $180^\circ$ out of phase as a function of pump-probe delay $\tau_p$. 
It is noted that Ref.~\cite{Stock1992} gives a theoretical treatment of the pump-probe signal in the transmission process for optically thin systems. Essentially, the same formula can be applied to the reflection process by regarding $E(t)$ as the electric field of the reflected probe pulse without the pump. 

In the commonly used transient-reflectivity measurements without filters, the oscillations at higher and lower frequencies cancel out because they oscillate $180^\circ$ out of phase. While one may think that no coherent phonon oscillation is observed without any filters, the oscillation amplitude is proportional to the difference between the electric susceptibilities, $\chi(\Omega)-\chi(\Omega \pm \omega)$, and the electric susceptibility is an increasing monotonic function in the transparent region encountered in Eq.~(\ref{final1}). In our case, the lower frequency components can be observed in the detection without a filter. The oscillation amplitude in $\Delta R(\Omega)/R(\Omega)$ with the specific filter is bigger than that in $\Delta R/R_0$ without a filter. Our experimental data show that the sensitivity of the FB900 band-pass filter is $35$ times higher than that without any filter. We remark that the observed oscillations with FB900 and FB850 are not perfectly $180^\circ$ out of phase. This may be caused by the spectral chirping in the optical pulse because the Fourier-transform-limited pulse was assumed for simplicity in our theoretical treatment. 

\section{Conclusion and Outlook}
We observed the changes in the transient reflectivity via coherent optical phonons in diamond using ultrashort infrared pulses and showed that the spectrally resolved detection can enhance the sensitivity of the coherent phonon measurement approximately $35$ times compared with the commonly used spectrally integrated measurement. This is because the reflected-light intensities for higher and lower frequencies from the spectral centroid  oscillate $180^\circ$ out of phase and canceled each other out.

With our enhanced method, we can measure the transient vibrational-state dynamics in a diamond more precisely, for example, the coherence time of the coherent phonon. As alluded to before~\cite{Lee2010,Lee2011,Lee2011B, England2, England}, the vibrational state of a diamond is a candidate for room-temperature based quantum memory. By utilizing coherent optical phonons in diamond, a coherent read-write process in quantum memory might be demonstrated using a pump-pump-probe method with wave-packet interference of optical phonons~\cite{Fleming, Ohmori}. 
\begin{acknowledgements}
The authors thank Yasuaki Okano for providing useful discussions and also thank Masaki Hada and Takuya Imaizumi for experimental helps. The authors thank Kyoko Kamo for help in the illustration of Fig.~\ref{expfig}. This work was partially supported by Core Research for Evolutional Science and Technology (CREST) of the Japan Science and Technology Agency (JST), JSPS KAKENHI Grant Numbers 25400330, 14J11318, 15K13377, and 16K05410, the Collaborative Research Project of Materials and Structures Laboratory, the DAIKO Foundation, and the Joint Studies Program of Institute for Molecular Science, National Institutes of Natural Sciences. Also, N. K. is partially supported by the Japan Society for the Promotion of Science.
\end{acknowledgements}

\appendix
\section{Detection mechanism of coherent phonons with spectral integration and resolution} \label{appendixA}
The whole process of generation and detection of the coherent phonons may be formulated as a series of 
higher-order optical processes. In this Appendix, we concentrate on the detection process, assuming that 
there exists a coherent lattice oscillation generated by a pump pulse irradiated in the past. 
The coherent phonon is regarded as a tool to apply an oscillating perturbation on the electrons in the crystal, 
and the probe pulse detects the change in the electronic states through the modulation of the frequency dispersed signal. 
It is noted that another quantum-mechanical treatment on the generation and detection mechanism of the coherent phonon was recently analyzed~\cite{Esposito}.

Let us consider a two-band system for the insulating crystal, say diamond, which describes the dynamics in the probe process. 
It is assumed that the energy of the excited states is modulated by the coherent oscillation of the LO phonon at the $\Gamma$-point, which has been generated by the pump pulse. The model Hamiltonian is given by
\begin{eqnarray}
H&=&H_g|g\rangle\langle g|+\sum_k H_k|k\rangle\langle k|,\nonumber\\
H_g&=&\hbar\omega b^\dagger b,\nonumber\\
H_k&=&\epsilon_k+\hbar\omega b^\dagger b-\alpha \hbar\omega \left(b+b^\dagger\right),
\end{eqnarray}
where the state vector $\ketg$ means the electronic ground state of the crystal, and $\ketk$ is the state for which an electron with wave vector $\vk$ is excited from $\ketg$ to the conduction band with the excitation energy $\epsilon_k$. The Hamiltonian $H_g$ and $H_k$ are the phonon Hamiltonians in the subspaces $\ketg$ and $\ketk$. The creation and the annihilation operators of the LO phonon at the $\Gamma$-point with energy $\hbar\omega$ are denoted by $b^\dagger$ and $b$, respectively. A similar model has been used to generate phononic states in quantum dots~\cite{Reiter2011}. It is assumed that the excited states are coupled with the LO phonon mode through the deformation potential interaction with the dimensionless coupling constant $\alpha $. In approximation, we have neglected the $\vk$-dependence of the coupling constant, assuming a rigid-band shift because of the deformation potential interaction. In the bulk crystal, the Huang--Rhys factor $\alpha^2$ is considered to be small, $\alpha^2 \ll 1$. 

Within the rotating wave approximation, the interaction Hamiltonian with the probe pulse is given by
\begin{equation}
H_I=\sum_k\mu_k E(t-\tau_p) |k\rangle\langle g|+ H. c.,
\label{int}
\end{equation}
in which $\mu_k$ is the transition dipole moment from $\ketg$ to $\ketk$, and $E(t - \tau_p)$ is the temporal profile 
of the electric field of the probe pulse with the delay time $\tau_p$. 
The spectral profile of the probe pulse is given by 
\begin{equation}
E(\Omega)=\int_{-\infty}^\infty E(t-\tau_p) e^{i\Omega t}dt=e^{i\Omega \tau_p}E_0(\Omega).
\end{equation} 
with
\begin{equation}
E_0(\Omega)=\int_{-\infty}^\infty E(t) e^{i\Omega t}dt.
\end{equation}
The interaction Hamiltonian (\ref{int}) can also be used also for the pump process by setting $\tau_p=0$. 
The only difference is that the amplitude of $E(t)$ is much larger than that in the probe pulse.

If one neglects the reflection from the back surface of the crystal, the pump-induced change of the reflection 
is obtained as a heterodyne modulation on the reflected probe pulse.  
Aside from irrelevant factors, the spectrally resolved signal of the reflection~\cite{Stock1992} is given by
\begin{equation}
\Delta R(\Omega)=2\Omega_0 {\rm Im}\left\{E^*(\Omega)\Delta P(\Omega)\right\},
\label{DeltaR}
\end{equation}
where $\Omega$ is the detection frequency and $\Delta P(\Omega)$ is the Fourier component 
of the modulated part of the polarization due to the coherent phonon,
\begin{equation}
\Delta P(\Omega)=\int_{-\infty}^\infty  \Delta P(t) e^{i\Omega t}dt.
\label{POmega}
\end{equation}
Equation (\ref{DeltaR}) can be derived from the argument that the small change in the reflected amplitude 
of the probe pulse originates from the loss or gain done by the work of the induced polarization in the presence 
of coherent lattice oscillation. 
It can be shown that the spectrally integrated signal, which is obtained in the common 
pump-probe experiment for the coherent phonons, is given by the integration of the spectrally resolved signal as 
\begin{eqnarray}
\Delta R&=&2\Omega_0\int_{-\infty}^\infty {\rm Im}\left\{E^*(t)\Delta P(t) \right\}\nonumber dt\\
&=&2\Omega_0\int_{-\infty}^\infty  {\rm Im}\left\{E^*(\Omega) \Delta P(\Omega) \right\}d\Omega.
\end{eqnarray}
The actually observed change in the reflection is normalized by the reflection amplitude $R(\Omega)$ 
without the pump pulse. As $R(\Omega)\propto |E(\Omega)|^2$, we find
\begin{equation}
\Delta R(\Omega)/R(\Omega)=2\Omega_0{\rm Im}\left\{\Delta P(\Omega)/E(\Omega)\right\}.
\end{equation}

For non-resonant excitation by the pump pulse, the coherent oscillation of the optical phonons is driven 
by the impulsive momentum generated by the virtual transition to the excited electronic states. 
We take the origin of time at the moment that the displacement of the phonon coordinate becomes maximum. 
Furthermore, we assume that the initial configuration of the phonon is given by the coherent state defined as
\begin{equation}
b|\beta\rangle = \beta|\beta\rangle, 
\end{equation}
for a real eigenvalue $\beta$. While this is an approximation, it makes the whole calculation transparent. 
Hence, the pump pulse prepares the initial state in which the electron is in the ground state while the phonon is 
in the excited state. For this initial state wave function 
\begin{equation}
|\psi(0)\rangle=|g\rangle \otimes |\beta\rangle,
\end{equation}
the Schr\"odinger equation
\begin{equation}
i\hbar\frac{d}{dt}|\psi(t)\rangle=\left(H+H_I\right)|\psi(t)\rangle
\end{equation}
is solved to the lowest order in $H_I$ as
\begin{widetext}
\begin{equation}
|\psi(t)\rangle =e^{-iH_g t/\hbar}|g\rangle\otimes |\beta\rangle 
+\frac{i}{\hbar}\int_0^t d\tau E(\tau-\tau_p)\sum_k \mu_k e^{-iH_k(t-\tau)/\hbar} 
e^{-iH_g\tau/\hbar}|k\rangle\otimes|\beta\rangle.
\label{psi}
\end{equation}
The density matrix for the electron-phonon system is given by $\rho(t)=|\psi(t)\rangle\langle \psi(t)|$. 
For the polarization operator $P^{op}\equiv \sum_k\mu_k^*|g\rangle\langle k|$, the complex polarization 
at $t$ is given by 
\begin{equation}
P(t)={\rm Tr}\left\{\rho(t)P^{op}\right\}.
\end{equation}

The third-order nonlinear polarization $\Delta P(t)$ comes from the cross term of the first and second terms of Eq.~(\ref{psi}). 
The time evolution and the expectation value of the phonon variable can be calculated by elementary arithmetic 
using the coherent-state representation. We obtain 
\begin{eqnarray}
\Delta P(t)&=& \frac{i}{\hbar} \int_0^t d\tau E(\tau-\tau_p)\sum_k|\mu_k|^2
e^{-i\epsilon_k(t-\tau)/\hbar-\gamma(t-\tau)} \nonumber\\
&\times& \exp\left[\alpha^2\{e^{-i\omega(t-\tau)}-1+i\omega (t-\tau)\}+2i\alpha\beta(\sin\omega t-\sin\omega \tau)\right]
\quad (t \geq 0),\nonumber\\
&=&0\quad (t<0),
\label{P}
\end{eqnarray}
with a small positive number $\gamma$ corresponding to the life time in the excited state. 
We are not concerned with the relaxation mechanism of the coherent phonon in the present work. 
The frequency-resolved polarization is given by Eq. (\ref{POmega}). 
For large enough $\tau_p$, $\Delta P(\Omega)$ is written as 
\begin{eqnarray}
\Delta P(\Omega)&=& \frac{i}{\hbar} \int_{-\infty}^\infty d\tau E(\tau-\tau_p)e^{i\Omega\tau}\int_0^\infty ds F(s) \nonumber\\
&\times& \exp\left[i\Omega s +\alpha^2(e^{-i\omega s} -1+i\omega s)
+2i\alpha\beta \left(\sin\omega(\tau+s)-\sin\omega\tau\right)\right],
\end{eqnarray}
where we set $t-\tau=s$ and the optical response function
\begin{eqnarray}
F(t)&=&\sum_k|\mu_k|^2e^{-i\epsilon_k t/\hbar-\gamma t},\quad \gamma=0_+ \quad (t\geq 0 )\nonumber\\
&=&0 \quad (t < 0).
\end{eqnarray}
The Fourier transform of $F(t)$ is the electric susceptibility of the material,
\begin{equation}
\chi(\Omega)=\frac{i}{\hbar}\int_{-\infty}^\infty F(t) e^{i\Omega t}dt .
\end{equation}
Applying the formula
\begin{equation}
e^{ix\sin\theta}=\sum_{m=-\infty}^\infty J_m(x)e^{im\theta}
\end{equation}
with $J_m(x)$ being the $m$th Bessel function, we find the final result in a closed expression,
\begin{eqnarray}
\Delta P(\Omega)&=&e^{-\alpha^2}\sum_{l=0}^\infty \sum_{m=-\infty}^\infty \sum_{n=-\infty}^\infty 
\frac{\alpha^{2l}}{l!}J_m(2\alpha\beta)J_n(2\alpha\beta)\nonumber\\
&\times& \chi(\Omega+(m-l)\omega+\alpha^2\omega)e^{i[\Omega+(m-n)\omega]\tau_p}
E_0(\Omega+m\omega-n\omega).
\end{eqnarray}
In bulk crystals, we may set $\alpha^2<1$ and $|\alpha\beta|<1$. In this case, we retain only terms $l=0$, $m=0,\pm 1$ and $n=0,\pm 1$. As a function of $\tau_p$, the oscillating part are given by terms $(m,n)=(\pm 1, 0)$ and $(0, \pm 1)$. As $\lim_{x\rightarrow 0_+}J_0(x)=1$, $\lim_{x\rightarrow 0+} J_1(x)=x/2$ and $J_{-1}(x)=-J_1(x)$, we find
\begin{eqnarray}
\Delta P(\Omega)&=&e^{i\Omega\tau_p}\alpha\beta [\left\{\chi(\Omega+\omega)
-\chi(\Omega)\right\}e^{i\omega\tau_p}E_0(\Omega+\omega)\nonumber\\
&+&\left\{\chi(\Omega)-\chi(\Omega-\omega)\right\}e^{-i\omega\tau_p}E_0(\Omega-\omega)].
\end{eqnarray}
Because the electric susceptibility is a real quantity in the transparent region, 
\begin{eqnarray}
\Delta R(\Omega)&=&2\Omega_0\alpha\beta [\left\{\chi(\Omega+\omega)
-\chi(\Omega)\right\}{\rm Im} \{ e^{i\omega\tau_p}E_0(\Omega+\omega)E^*_0(\Omega) \} \nonumber\\
&+&\left\{\chi(\Omega)
-\chi(\Omega-\omega)\right\}{\rm Im} \{ e^{-i\omega\tau_p}E_0(\Omega-\omega)E^*_0(\Omega) \} ].
\end{eqnarray}
If the probe pulse is in the Fourier transform limit, $E_0(\Omega)$ is also a real quantity  
and the above formula reduces to a much simpler expression, 
\begin{eqnarray}
\Delta R(\Omega)&=&2\Omega_0 \alpha\beta [\left\{\chi(\Omega+\omega)-\chi(\Omega)\right\}
E_0(\Omega+\omega)
\nonumber\\
&-&\left\{\chi(\Omega)-\chi(\Omega-\omega)\right\}E (\Omega-\omega)
]  E_0(\Omega)\sin\omega\tau_p.
\label{final}
\end{eqnarray}
\end{widetext}
If one introduces the coordinate of the phonon as 
\begin{equation}
Q\equiv \sqrt{\hbar/2\omega}(b+b^\dagger),
\end{equation}
the amplitude of oscillation $Q_0$ is given by 
\begin{equation}
Q_0=\langle \beta|Q|\beta \rangle =\sqrt{2\hbar/\omega}\beta.
\end{equation}
This recovers Eq. (\ref{final1}).

For a frequency at the peak position of the incident pulse $\Omega_0$, the peak frequency in the reflected signal becomes $\Omega_0-\omega$ in the first term in Eq.~(\ref{final}), and $\Omega_1+\omega$ in the second term. The phase of the oscillation is reversed in both the low-frequency and the high-frequency parts. Furthermore, the oscillation amplitude at the high-frequency peak is slightly larger than the lower one if the pulse is in the Fourier transform limit because the second derivative of the susceptibility is usually positive in the transparent region;
\begin{equation}
\frac{\partial^2\chi}{\partial \Omega^2} >0
\end{equation}
so that  
\begin{equation}
|\chi(\Omega+\omega)-\chi(\Omega)|>|\chi(\Omega)-\chi(\Omega-\omega)|.
\end{equation}
Therefore, the change in transient reflectivity is not completely canceled out and can be measured in the commonly used spectrally integrated measurement.


\begin{thebibliography}{99}
\bibitem{Dekorsy2000} T. Dekorsy, G. C. Cho, and H. Kurz, in \textit{Light Scattering in Solids VIII}, (eds.) M. Cardona and G. G{\"u}ntherodt, (Springer, Berlin, 2000) pp. 169--209. 
\bibitem{Nelson1985} Y.-X. Yan, E. B. Gamble, and K. Nelson, J. Chem. Phys. \textbf{83}, 5391 (1985). 
\bibitem{Cheng1991} T. K. Cheng, J. Vidal, H. J. Zeiger, G. Dresselhaus, M. S. Dresselhaus, and E. P. Ippen, Appl. Phys. Lett. \textbf{59}, 1923 (1991). 
\bibitem{Zeiger1992} H. J. Zeiger, J. Vidal, T.K. Cheng, E. P. Ippen, G. Dresselhaus and M.S. Dresselhaus, Phys. Rev. B \textbf{45}, 768 (1992).
\bibitem{DeCamp2001} M. F. DeCamp, D. A. Reis, P. H. Bucksbaum, and R. Merlin, Phys. Rev. B \textbf{64}, 092301 (2001). 
\bibitem{Katsuki2013} H. Katsuki, J. C. Delagnes, K. Hosaka, K. Ishioka, H. Chiba, E. S. Zijlstra, M. E. Garcia, H. 
Takahashi, K. Watanabe, M. Kitajima, Y. Matsumoto, K. G. Nakamura, and K. Ohmori, Nat. Commun. \textbf{4}, 2801 (2013).
\bibitem{Cho1990} G. C. Cho, W. K\"utt, and H. Kurz, Phys. Rev. Lett. \textbf{65}, 764 (1990).
\bibitem{Dekorsy1993} T. Dekorsy, T. Pfeifer, W. K\"utt, and H. Kurz, Phys. Rev. B \textbf{47}, 3842 (1993).
\bibitem{Merlin1996} G. A. Garrett, T. F. Albrecht, J. F. Whitaker and R. Merlin, Phys. Rev. Lett., \textbf{77}, 3661 (1996).
\bibitem{Misochko2000} O. V. Misochko, K. Kisoda, K. Sakai, and S. Nakashima, Phys. Rev. B {\bf 61}, 4305 (2000).
\bibitem{Hase2003} M. Hase, M. Kitajima, A. M. Constantinescu, and H. Petek, Nature \textbf{426}, 51 (2003).
\bibitem{Hayashi2014} S. Hayashi, K. Kato, K. Norimatsu, M. Hada, Y. Kayanuma, and K. G. Nakamura, Sci. Rep. \textbf{4}, 4456 (2014).
\bibitem{Chwalek1991} J. M. Chwalek, C. Uher, J. F. Whitaker, G. A. Mourou, and J. A. Agostinelli, Appl. Phys. Lett. \textbf{58}, 980 (1991). 
\bibitem{Albrecht1991} W. Albrecht, Th. Kruse, and H. Kurz, Phys. Rev. Lett. {\bf 69}, 1451 (1992).
\bibitem{Takahashi2011} H. Takahashi, Y. Kamihara, H. Koguchi, T. Atou, H. Hosono, I. Katayama, J. Takeda, M. Kitajima, and K. G. Nakamura, J. Phys. Soc. Jpn. {\bf 80}, 013707 (2011).
\bibitem{Wu2008} A. Q. Wu, X. Xu, and R. Venkatasubramanian, Appl. Phys. Lett. \textbf{92}, 011108 (2008). 
\bibitem{Kamaraju2010} N. Kamaraju, S. Kumar, and A. K. Sood, Europhys. Lett. \textbf{92}, 47007 (2010). 
\bibitem{Norimatsu2014} K. Norimatsu, J. Hu, A. Goto, K. Igarashi, T. Sasagawa, and K. G. Nakamura, Solid State Commun. \textbf{157}, 58 (2013).
\bibitem{Misochko2015} O. V. Misochko, J. Flock, and T. Dekorsy, Phys. Rev. B \textbf{91}, 174303 (2015). 
\bibitem{Ishioka2008GR} K. Ishioka, M. Hase, M. Kitajima, L. Wirtz, A. Rubio, and H. Petek, Phys. Rev. B \textbf{77}, 121402(R) (2008). 
\bibitem{Katayama2013} I. Katayama, K. Sato, S. Koga, J. Takeda, S. Hishita, H. Fukidome, M. Suemitsu, and M. Kitajima, Phys. Rev. B, \textbf{88}, 245406 (2013).  
\bibitem{Kim2013} J.-H. Kim, A.R.T. Nugraha, L.G. Booshehri, E.H. H\'aroz, K. Sato, G. D. Sanders, K.-J. Yee, Y.-S. Lim, C. J. Stanton, R. Saito, and J. Kono, Chem. Phys. \textbf{413}, 55 (2013). 
\bibitem{Kato2008} K. Kato, K. Ishioka, M. Kitajima, J. Tang, R. Saito, and H. Petek, Nano Lett. \textbf{8}, 3102 (2008). 
\bibitem{Gambetta2006} A. Gambetta, C. Manzoni, E. Menna, M. Meneghetti, G. Cerullo, G. Lanzani, S. Tretiak, A. Piryatinski, A. Saxena, R.L. Martin, and A. R. Bishop, Nat. Phys. \textbf{2}, 515 (2006). 
\bibitem{Sanders2009} G. D. Sanders, C. J. Stanton, J.-H. Kim, K.-J. Yee, Y.-S. Lim, E. H. H\'aroz, L. G. Booshehri,  J. Kono, and R. Saito, Phys. Rev. B \textbf{79}, 205434 (2009). 
\bibitem{Makino2009} K. Makino, A. Hirano, K. Shiraki, Y. Maeda, and M. Hase, Phys. Rev. B \textbf{80}, 245428 (2009). 
\bibitem{Kim2009} J.-H. Kim, K.-J. Han, N.-J. Kim, K.-J. Yee, Y.-S. Lim, G. D. Sanders, C. J. Stanton, L. G. Booshehri, E. H. H\'aroz, and J. Kono, Phys. Rev. Lett. \textbf{102}, 037402 (2009).  
\bibitem{Merlin1997}  R. Merlin, Solid State Commun. \textbf{102}, 207 (1997). 
\bibitem{Misochko2004} O. V. Misochoko, M. Hase, and M. Kitajima, J. Phys.: Condens. Matter \textbf{16}, 1879 (2004).  
\bibitem{Mizoguchi2013} K. Mizoguchi, R. Morishita, and G. Oohata, Phys. Rev. Lett. \textbf{110}, 077402 (2013). 
\bibitem{Lee2010} K. C. Lee, B. J. Sussman, J. Nunn, V. O. Lorenz, K. Reim, D. Jaksch, I. A. Walmsley, P. Spizzirri, and S. Prawer,  Diam. Rel. Mater,  \textbf{19}, 1289 (2010). 
\bibitem{Lee2011} K. C. Lee, M. R. Sprague, B. J. Sussman, J. Nunn, N. K. Langford, X.-M. Jin, T. Champion, P. Michelberger, K. F. Reim, D. England, D. Jaksch, I. A. Walmsley, Science \textbf{334}, 1253 (2011).
\bibitem{Lee2011B} K. C. Lee, B. J. Sussman, M. R. Sprague, P. Michelberger, K. F. Reim, J. Nunn, N. K. Langford, P. J. Bustard, D. Jaksch and I. A. Walmsley, Nat. Photon. \textbf{6}, 41 (2012).
\bibitem{England2} D. G. England, P. J. Bustard, J. Nunn, R. Lausten, and B. J. Sussman, Phys. Rev. Lett. \textbf{111}, 243601 (2013).
\bibitem{England} D. G. England, K. A. G. Fisher, J.-P. W. MacLean, P. J. Bustard, R. Lausten, K. J. Resch, and B. J. Sussman, Phys. Rev. Lett. \textbf{114}, 053602 (2015).
\bibitem{Ishioka2006} K. Ishioka, M. Hase, M. Kitajima, and H. Petek, Appl. Phys. Lett. \textbf{89}, 231916 (2006). 
\bibitem{Nakamura2015} K. G. Nakamura, Y. Shikano, and Y. Kayanuma, Phys. Rev. B \textbf{92}, 144304 (2015).
\bibitem{Milden2013} R. P. Milden, in {\it Optical Engineering of Diamond}, (eds.) R. P. Milden and J. R. Rabeau (Wiley-VCH Verlag \& Co. KGaA, Weinheim, 2013), pp. 1--34.
\bibitem{Stock1992}  G. Stock and W. Domcke, Phys. Rev. A \textbf{45}, 3032 (1992). 
\bibitem{Fleming} S. J. Rosenthal, X. Xie, M. Du, and G. R. Fleming, J. Chem. Phys. {\bf 95}, 4715 (1991). 
\bibitem{Ohmori} K. Ohmori, Annu. Rev. Phys. Chem. {\bf 60}, 487 (2009).
\bibitem{Esposito} M. Esposito, K. Titimbo, K. Zimmermann, F. Giusti, F. Randi, D. Boschetto, F. Parmigiani, R. Floreanini, F. Benatti, D. Fausti, Nature Comm. {\bf 6}, 10249 (2015).
\bibitem{Reiter2011} D. E. Reiter, D. Wigger, V. M. Axt, and T. Kuhn, Phys. Rev. B \textbf{84}, 195327 (2011).  
\end{thebibliography}
\end{document}